# Inversion of a two-level atom by quantum superoscillations

I. V. Doronin[1], A. A. Pukhov[1,2,3], E. S. Andrianov[1,2], A. P. Vinogradov[1,2,3], and A. A. Lisyansky[4,5]

[1]Moscow Institute of Physics and Technology, 9 Institutskiy per., 141700 Dolgoprudny, Moscow reg., Russia
[2]Dukhov Research Institute for Automatics, 22 Sushchevskaya, Moscow 127055, Russia
[3]Institute for Theoretical and Applied Electromagnetics RAS, 13 Izhorskaya, Moscow 125412, Russia
[4]Department of Physics, Queens College of the City University of New York, Queens, NY 11367, USA
[5]The Graduate Center of the City University of New York, New York, New York 10016, USA

We show that a two-level atom with a high transition frequency $\omega_{SO}$ can be inverted via non-radiative interaction with a cluster of excited low frequency two-level atoms or quantum oscillators whose transition frequencies are smaller than $\omega_{SO}$. This phenomenon occurs due to the Förster resonant energy transfer arising during a train of quantum superoscillation of low frequency two-level atoms. The suggested model could explain the mechanism of biophoton emission.

Mitogenetic rays called biophotons is weak (about 300-1,000 photons per second per square centimeter) radiation [1-3] emitted by living organisms during mitosis [4]. Biophotons were detected in optical and near ultraviolet ranges for various biological systems [5]. Biophotons may play an important role in processes of cell division and are proposed to have applications in medical diagnostics [4].

The mechanism of the biophoton generation is still unknown. In complex organic molecules, only vibronic and conformational degrees of freedom may be excited at room temperature. The transition frequencies corresponding to such excitations lie in the infrared range. It is not clear how these low-energy quanta may excite a high-energy electron state to radiate biophotons.

In our communication, we examine quantum superoscillations (SO) [6-8] as a possible mechanism for biophoton excitations. A complex system of levels in organic molecules is modeled by a system of two-level atoms (TLA's) and/or quantum oscillators. In particular, we study a quantum problem in which a TLA with a high transition frequency is excited by the energy transferred from a cluster of low-frequency (LF) TLA's or quantum oscillators. We show that even though an energy quantum of each LF TLA is lower than the energy required for the excitation of the high-frequency (HF) TLA near fields of all LF TLA's allows for the inversion of the HF TLA. The price paid for superoscillating behavior is that the total energy of LF oscillations must substantially exceed the excitation energy of the HF TLA.

The phenomenon of SO is a counterintuitive mathematical effect showing that during some finite time interval, a sum of band-limited functions can oscillate with a frequency that is higher than the maximum frequency of their spectra [9, 10]. This phenomenon is widely used in such fields as quantum physics [6-8], optics [11-14], and radiolocation [15-17].

Let us consider a HF TLA with a transition frequency of $\omega_{SO}$ and $n$ clusters with $N$ LF TLA's in each with frequencies $\omega_1,...,\omega_n \leq 0.9\omega_{SO}$. We assume that all participating particles are confined to a subwavelength volume. Consequently, the energy is mainly transferred nonradiatively, by near fields. For simplicity, the interaction of LF TLA's with each other is neglected.

The system is described by the dipole-dipole interaction Hamiltonian [18]:

$$H = \sum_{j=1}^{n}\sum_{i=1}^{N}\hbar\omega_j\hat{\sigma}_{ij}^+\hat{\sigma}_{ij} + \hbar\omega_{SO}\hat{\sigma}^+\hat{\sigma} + \hbar(\hat{\sigma}+\hat{\sigma}^+)\hat{F}(t), \qquad (1)$$

where $\hat{\sigma}^+$ and $\hat{\sigma}$ are operators of excitation and de-excitation of the HF TLA, $\hat{\sigma}_{ij}^+$ and $\hat{\sigma}_{ij}$ are operators of excitation and de-excitation of the $i$-th LF TLA from the $j$-th cluster, $\hat{F}(t) = \sum_{j=1}^{n}\sum_{i=1}^{N}\Omega_{ij}\left(\hat{\sigma}_{ij}(t)+\hat{\sigma}_{ij}^+(t)\right)$ is the Förster driving operator which describes the effect of local fields created by LF TLA's on the HF TLA, $\Omega_{ij} = \mathbf{d}\cdot\mathbf{d}_{ij}/\left(|\mathbf{r}-\mathbf{r}_{ij}|\right)^3$ ($j=1,...,n$, $i=1,...,N$) are coupling constants of the dipole moment $\mathbf{d}_{ij}(\hat{\sigma}_{ij}(t)+\hat{\sigma}_{ij}^+(t))$ of $i$-th LF TLA from the $j$-th cluster with the dipole moment $\mathbf{d}(\hat{\sigma}(t)+\hat{\sigma}^+(t))$ of the HF TLA, $\mathbf{d}$ and $\mathbf{d}_{ij}$ are the matrix elements of dipole moments of the HF and a LF TLA's, $\mathbf{r}$ and $\mathbf{r}_{ij}$ are the positions of the HF TLA and the $j$-th LF TLA, respectively. The inversion operator $\hat{\sigma}_{ij}^+\hat{\sigma}_{ij} - \hat{\sigma}_{ij}\hat{\sigma}_{ij}^+$ of the $i$-th LF TLA from the $j$-th cluster we denote as $\hat{D}_{ij}$, while $\hat{D}$ is the inversion operator of the HF TLA.

The Hamiltonian (1) does not contain terms describing electromagnetic modes (photons). Nonetheless, between TLA's, the energy can be transferred by near fields via the Förster mechanism [19-21]. At distances smaller than the wavelength, this energy is by $(kr)^3$ greater than that of far-fields. The energy transfer by near fields has a resonance nature [22]. For example, when the frequency detuning of two interacting TLA's is 10%, the population inversion of HF TLA stays below $-0.8$ [18].

First, we estimate characteristic parameters of the quantities that we deal with. The value of the dipole moment can reach 150 D (for Rhodamine 800, it is 145 D [23]), we take distances between TLA's as ~20 nm, then $\hbar\Omega \sim d^2/r^3 \sim 10^{-14}$ erg, and therefore, $\Omega \sim 3.0 \cdot 10^{13}$ s$^{-1}$, in the optical range, the characteristic frequency $\omega_0 \sim 10^{15}$ s$^{-1}$. Below we use dimensionless time and frequency, $t \to t\omega_0$ and $\omega \to \omega/\omega_0$.

To describe the dynamics of the system, we use the approach based on the optical Bloch equations. First, we obtain the Heisenberg equations of motion for operators $\hat{\sigma}, \hat{\sigma}_{ij}, \hat{D}$, and $\hat{D}_{ij}$. Then, we move from operators to their expectation values $<\hat{\sigma}>=\sigma, <\hat{\sigma}_{ij}>=\sigma_{ij}, <\hat{D}>=D$, and $<\hat{D}_{ij}>=D_{ij}$. The resulting equations contain higher order correlators for which new equations should be obtained. To terminate an infinite chain of equations, we decouple correlators of the second order as $<\hat{\sigma}\hat{\sigma}_{ij}>=\sigma\sigma_{ij}$. Below, we consider a system of four clusters with $N = 10$ LF TLA's in each. For the sake of simplicity, we assume that in each cluster, all TLA's are the same.

As a result, we obtain a system of equations for *c*-numbers:

$$\dot{\sigma} = -i\omega_{SO}\sigma + iDN\sum_{j=1}^{n}\Omega_j\left(\sigma_j + \sigma_j^*\right) - \gamma\sigma,$$
$$\dot{D} = 2i\left(\sigma - \sigma^*\right)N\sum_{j=1}^{n}\Omega_j\left(\sigma_j + \sigma_j^*\right) - 2\gamma D,$$
$$\dot{\sigma}_j = -i\omega_j\sigma_j + i\Omega_j D_j\left(\sigma + \sigma^*\right) - \gamma_j\sigma_j, \quad (2)$$
$$\dot{D}_j = 2i\Omega_j\left(\sigma + \sigma^*\right)\left(\sigma_j - \sigma_j^*\right) - 2\gamma_j D_j,$$

where $\omega_j$ and $\Omega_j$ are dimensionless and $\omega_{SO} = 1$. In each equation, the last term describes attenuation in the $\tau$-approximation, the corresponding relaxation rate is denoted as $\gamma$. For numerical calculations we set $\gamma = \gamma_j = 0.001 << \omega_j$.

The dynamics of HF TLA driven by the operator $\hat{F}(t)$, which plays the role of an external field, shows complex Rabi oscillations [24]. To obtain a noticeable inversion, $\hat{F}(t)$ should resonantly affect the HF TLA transiton. This may occure during the SO train. Indeed if the relation $\Omega_F \Delta t \approx \pi$ is fulfilled, the inversion should achieve its maximum value due to the Rabi oscillations. Here, $\Delta t$ is a duration of the SO train, $\Omega_F$ is the effective Rabi frequency $\left(2(\Delta t)^{-1}\int_{\Delta t}F^2(t)dt\right)^{1/2}$ in the Förster field. Since, an increase of the SO train duration usually leads to a decrease in the amplitude of the HF field, which in turn, decreases the Rabi frequency, the value of $\Omega_F$ must be sufficiently large. This requires a large expectation value $F(t)$ of the operator $\hat{F}(t)$.

A direct way of increasing the Rabi frequency is by raising the coupling constants $\Omega_j$. However, in our calculations, we have used maximal values of $\Omega_j = 0.01$ known from experiment [23]. Their further increase is not realistic. A larger Rabi frequency can be achieved by increasing the number of TLA clusters and the number of TLA's in each cluster.

We optimize the objective function $\langle D\rangle_{50-300} = \int_{50}^{300}D(t)dt/250$ assuming that the frequency of LF TLA's vary within the limit: $0 < \omega_1,...,\omega_4 < 0.9$. The optimization gives $\langle D\rangle_{50-300} = 0.485$ for the following values of the frequencies: $\omega_1 = 0.36$, $\omega_2 = 0.55$, $\omega_3 = 0.73$, $\omega_4 = 0.88$, and initial conditions $\arg(\sigma_1) = 0.06$, $D_1(0) = 0.04$, $\arg(\sigma_2) = 0.20$, $D_2(0) = 0.00$, $\arg(\sigma_3) = 5.89$, $D_3(0) = 0.17$, $\arg(\sigma_4) = 4.73$, and $D_4(0) = 0.08$.

Even though the maximum initial energy in the system is achieved when the population inversion of LF TLA's is $D = 1$, the optimized values of $D$ are rather small. This happens due to the necessity of increasing the Rabi frequency, which is proportional to the field amplitude. The maximum values of fields of LF TLA's are proportional to $\sum_j \Omega_j$ and are achieved for $D_j(0) = 0$. Indeed, $D$ and $|\sigma|$ are interrelated. For a pure TLA state, we have $|\psi\rangle = C_e |e\rangle + C_g |g\rangle$, thus $\sigma = C_g^* C_e$ and $D = |C_e|^2 - |C_g|^2$ by definition [18]. Using the condition $|C_e|^2 + |C_g|^2 = 1$, we obtain $1 - D = 2|C_g|^2$ and similarly $1 + D = 2|C_e|^2$. This gives $1 - D^2 = 4|C_g^* C_e|^2 = 4|\sigma|^2$. Thus, the maximum amplitude of near-fields is reached for $D_j(0) = 0$; then $|\sigma_j(0)| = 0.5$.

The dynamics of the system with the parameters obtained during optimization is shown in Fig. 1. The calculations show that there are several trains of SO. The first train lasts from $t = 0$ to $t = 10\pi$ (Fig 2). Its spectrum is shown in Fig. 3. As we can see, the harmonic with the frequency $\omega_{SO} = 1$ dominates. For comparison, we present a spectrum for a time interval without SO (Fig. 3). Thus, during an SO train, the population inversion of the HF TLA should oscillate with the Rabi frequency $\Omega_F$, which is equal to $\approx 0.15$ for the first SO train.

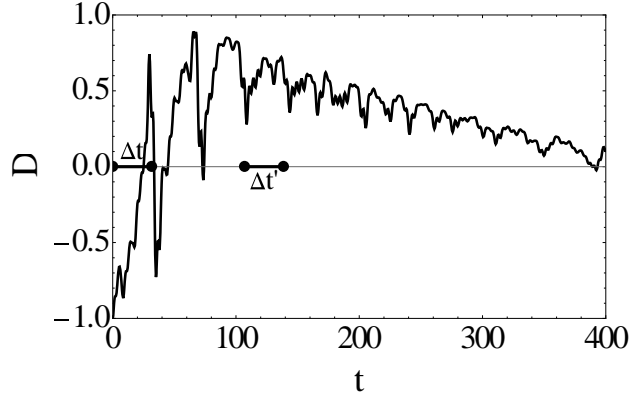

Fig. 1. The dynamics of the population inversion of the HF TLA. The interval $\Delta t$ corresponds to the SO train, the interval $\Delta t'$ is a typical interval without SO.

As one can see from Fig. 2, the excitation time $\Delta t_{exc}$ of the HF TLA approximately coincides with the duration of the SO train, $\Delta t = 10\pi$, and is approximately equal to five periods of the SO. The product $\Omega_F \Delta t_{exc}$ is $\approx 4.75$ which is in good agreement with the estimation $\Omega_F \Delta t_{exc} \approx \pi$ following from the theory of the Rabi oscillations [18].

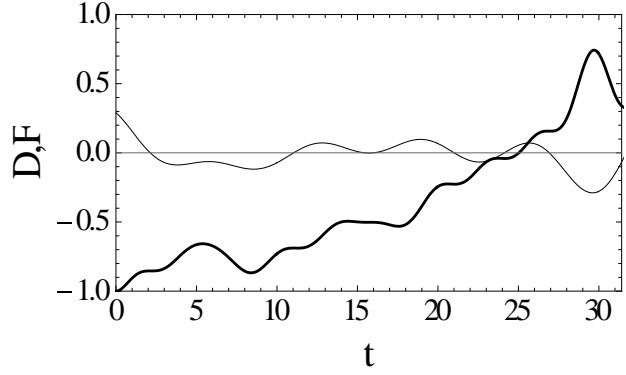

Fig. 2. The dynamics of the system during the first train of SO. The thick line shows the population inversion of the HF TLA; the thin line corresponds to the field amplitude created by LF TLA's at the location of the HF TLA. $|F| \leq \sum_{j=1}^{n}\sum_{i=1}^{N} 0.02|\sigma_{ij}(t)| \leq 0.4$.

The HF TLA is being excited until the train of the SO lasts. When the SO ends, the Rabi oscillations stop until the next SO train. After all trains end, the HF TLA remains in the excited state during the time which is much larger than the inverse Rabi frequency (Fig. 1).

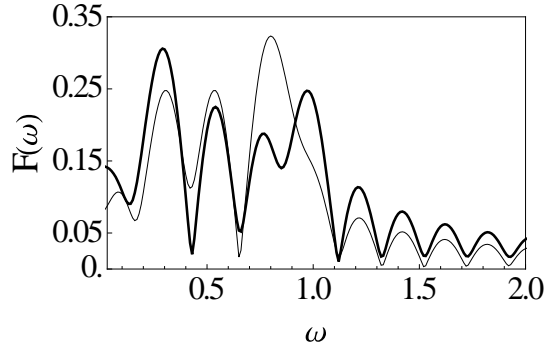

Fig. 3. Fourier spectra of the field amplitude $F$ for the train of SO $\Delta t = [0, 10\pi]$ (the thick line) and for the time interval without SO, $\Delta t' = [34\pi, 44\pi]$ (the thin line).

The phenomenon considered can be generalized for the interaction of the HF TLA with quantum harmonic oscillators, i.e. with multi-level systems. The Hamiltonian for such a problem can be written as

$$H = \sum_{j=1}^{n} \hbar\omega_j \hat{a}_j^+ \hat{a}_j + \hbar\omega_{SO}\hat{\sigma}^+\hat{\sigma} + \hbar(\hat{\sigma}+\hat{\sigma}^+)\hat{F}(t), \tag{3}$$

where $\hat{a}_j$ and $\hat{a}_j^+$ are operators of annihilation and creation of an excitation of the $j$-th quantum oscillator, $\Omega_j$ are coupling constants of the $j$-th oscillator with the HF TLA and $\hat{F}(t) = \sum_{j=1}^{n}\Omega_j(\hat{a}_j(t)+\hat{a}_j^+(t))$ is the corresponding Förster driving operator. To describe this system, one can obtain a system of equations for expectation values $<\hat{D}>=D$, $<\hat{\sigma}>=\sigma$, and $<\hat{a}_j>=a$. Similar to system of equations (3) we have

$$\dot{\sigma} = -i\omega_{SO}\sigma + iD\sum_{i=1}^{n}\Omega_i\left(a_i + a_i^*\right) - \gamma D,$$
$$\dot{D} = 2i\left(\sigma - \sigma^*\right)\sum_{i=1}^{n}\Omega_i\left(a_i + a_i^*\right) - 2\gamma D, \quad (4)$$
$$\dot{a}_j = -i\omega_j a_j - i\Omega_j\left(\sigma + \sigma^*\right) - \gamma_{aj}a_j.$$

Equations (4) for four quantum oscillators with initial conditions similar to the system of TLA's described above give the dynamics of the population inversion of the HF TLA very similar to the one shown in Fig. 1.

If the initial state of the system is not an eigenstate of the Hamiltonian, then the energy of the system is not defined. There is an energy spread due to which the energy may be exchanged between LF and HF TLA's. For the inversion $<\hat{D}_{ij}>\approx 0$, the dimensionless energy spread can be estimated as

$$\left\langle\left(\sum_{ij}\omega_{ij}\hat{D}_{ij}\right)^2\right\rangle - \left\langle\sum_{ij}\omega_{ij}\hat{D}_{ij}\right\rangle^2 \approx \left\langle\left(\sum_{ij}\omega_{ij}\hat{D}_{ij}\right)^2\right\rangle. \quad (5)$$

As in previous sections, $\omega_{ij}$ are dimensionless. Since averages of cross-terms are negligible, we have

$$\left\langle\left(\sum_{ij}\omega_{ij}\hat{D}_{ij}\right)^2\right\rangle \approx \sum_{ij}\omega_{ij}^2\left\langle\hat{D}_{ij}^2\right\rangle. \quad (6)$$

Taking into account that, by definition, $\hat{D}_{ij}^2 \equiv \hat{1}$, for the energy spread we obtain

$$\left\langle\sum_{ij}\left(\omega_{ij}\hat{D}_{ij}\right)^2\right\rangle = \sum_{ij}\left(\omega_{ij}\right)^2\left\langle\hat{D}_{ij}\right\rangle^2 = \sum_{ij}\left(\omega_{ij}\right)^2. \quad (7)$$

Substituting the values given by the optimization into Eq. (7) we obtain $\sum_{ij}\omega_{ij}^2 \approx 17\omega_{SO}^2 \gg \omega_{SO}^2$. Thus, the energy dispersion allows for the excitation of the HF TLA. Now, let us estimate the probability of such an excitation. For this purpose, we calculate the objective function for initial phases of near-fields from the interval $[-\pi,\pi]$ with the step of $\pi/5$ and initial population inversions from the interval [–0.9, 0.9] with the step of 0.1. Altogether, we obtain 190,000 system configurations and calculate the distribution $p_f = df/d<D>_{50-300}$, where $f\left(<D>_{50-300}\right)$ is a number of the systems in which the value of an averaged inversion population smaller than $<D>_{50-300}$ is realized. High values of population inversions are observed in very rare cases. This means that the probability of observing SO is less than $10^{-5}$. The maximum population inversion is in the interval [0.48, 0.50], where SO are observed. The solution obtained above, $<D>_{50-300} = 0.49$, belongs to this interval.

We show that a HF TLA can be inverted by near-fields of spontaneously relaxing LF TLA's. This phenomenon occurs due to an excitation of the resonant Rabi oscillations of the HF TLA by a train of SO of the driving operator. This is demonstrated for a system in which the HF TLA is excited at the instant when the HF packet is terminated. This means that the relaxation of the HF TLA into the lower state is not resonant, i.e. it does not happen during the Rabi period, but due to the spontaneous radiation of a photon during the lifetime of the upper state.

We have generalized the observed phenomenon for the interaction of the HF TLA with quantum oscillators. Thus, the energy can be transferred to the HF TLA from electron, rotational and vibrational transitions. Therefore, the discussed phenomenon can be a good candidate for the mechanism causing biophotons radiation. Of course, as we estimate, the probability of such a process is extremely small, but it can be realized and fortified in biological systems which evolutionary selection has taken billions of years.

**Funding.** National Science Foundation (NSF) (DMR-1312707).